\documentclass[aps,prb,preprint]{revtex4-1}
\usepackage{amsmath,amssymb,amsfonts,graphicx}
\usepackage{hyperref}
\begin{document}

\title{Comments on Kerr effect and  gyrotropic order in cuprates}

\author{Sudip Chakravarty}
\email{sudip@physics.ucla.edu}
\affiliation{Department of Physics and Astronomy, University of California Los Angeles, Los Angeles, California 90095, USA}

\date{\today}

\begin{abstract}
I comment on two recent papers on Kerr effect as evidence of gyrotropic order in cuprates, and I suggest that the arguments
may not be sound. The  difficulty is that in practically all cases the wave vector $k_{z}$ perpendicular to the copper-oxygen plane is not a 
good quantum number. This appears to be problematic for arXiv:1212.2698, whereas in arXiv:1212.2274 the symmetry arguments may turn out to be robust, but the
microscopic picture is wanting. Thus, the Kerr effect in cuprates remains a puzzle, as there is little doubt that the arguments presented against 
time reversal symmetry breaking appear to be rather strong in both of these  papers on  experimental grounds.
\end{abstract}
\maketitle
An extensive set of measurements have found polar Kerr effect in a large number of cuprate superconductors  at an  onset temperature
in the pseudogap regime; the experiments   are surveyed in a recent article.\cite{Kapitulnik:2011} Initially, the experiments were interpreted in terms of time reversal
symmetry breaking. However, an alternate proposal argues in favor of  spontaneous breaking of inversion and mirror reflection symmetries.~\cite{Hosur:2012,Orenstein:2012}
Both papers invoke gyrotropy of the medium~\cite{Landau:1984} in  the pseudogap state of the cuprates. The arguments against time reversal symmetry breaking have been  elegantly summarized, so there is little reason to duplicate them here. Here I want to question a fundamental assumption made in Ref.~\onlinecite{Orenstein:2012}, which is also relevant for Ref.~\onlinecite{Hosur:2012}. However,  the   symmetry based arguments and a phenomenological model discussed in Ref.~\onlinecite{Hosur:2012} could be  robust. 

The starting point of Ref.~\onlinecite{Orenstein:2012} is the equation describing the velocity,  ${\bf v}({\bf k})$, where it is augmented correctly by a so-called anomalous term~\cite{Karplus:1954} containing the Berry curvature ${\bf \Omega}(\bf k)$, 
\begin{equation}
{\bf v}({\bf k})=\frac{1}{\hbar}\frac{\partial \varepsilon({\bf k})}{\partial {\bf k}} - \frac{e}{\hbar} {\bf E}\times {\bf \Omega}(\bf k),
\end{equation}
where $\bf E$ is the electric field. Following Ref.~\onlinecite{Landau:1984} they then arrive at the expression for the maximum value of the Kerr rotation angle $\theta_{K}$ at normal incidence on the optic axis of a uniaxial crystal, which is\begin{equation}
\theta_{K} \sim \alpha l_{z} \Re \left[\frac{1}{(1-i\omega \tau_{z})^{2}[\epsilon_{\parallel}(\omega)-1]}\right] ,
\end{equation}
where $\epsilon_{\parallel}(\omega)$ is the dielectric function in the plane perpendicular to the optic axis  and $l_{z}$ is the mean free path in the $z$-direction, and $\alpha$ is the fine structure constant. This equation presupposes an order of magnitude estimate of the Berry phase contribution,~\cite{footnote1}
\begin{equation}
\int_{-k_{Fz}}^{k_{Fz}}dk_{z}v_{z}(k_{z})\int_{0}^{k_{F}(k_{z})} d^{2}k\;  \Omega_{z}(k,k_{z}) \sim 1.
\end{equation}
The entire derivation is based on $k_{z}$ being a good quantum number, which it is not for any of the under doped cuprates. Therefore, the efficacy of the formula for $\theta_{K}$ is moot. The authors then go on to make an estimate for the Kerr data in $\mathrm{La_{2-x}Ba_{x}CuO_{4}}$  by setting $l_{z}\sim  1$  (that is, the $c$-axis lattice constant), recognizing themselves that the $c$-axis transport is incoherent. For incoherent transport, the semiclassical dynamics involving $k_{z}$, whether or not one includes the Berry curvature, is not particularly meaningful.

A microscopic Hamiltonian which is explicitly coherent in interlayer tunneling in  Ref.~\onlinecite{Hosur:2012} appears to be problematic. Take, for example, their Eq. (9),  where the gyrotropic  tensor 
is expressed as 
\begin{equation}
\gamma(\omega)\propto \left(\frac{t_{\perp}}{E_{F}}\right)^{2} \cdots
\end{equation}
with no other dependences on the interlayer hopping matrix element $t_{\perp}$ (assumed to be momentum independent) and $E_{F}$ the Fermi energy. It is difficult to see how this equation will survive if incoherence in the direction perpendicular to the copper oxide plane is included. In any case, a well defined cholesteric  pitch requires coherent $c$-axis tunneling for which experimental evidence is non existent in any generic underdoped cuprates, some of which were investigated in Ref.~\onlinecite{Kapitulnik:2011}.

A few brief comments regarding incoherent interlayer hopping are in order. At temperatures above the superconducting transition temperature even the best underdoped  cuprate superconductors show an insulating upturn as the temperature is lowered with the  $c$-axis resistivity of the order of Ohm-cm or greater. In fact $c$-axis incoherence was used to explain quantum oscillations in $\rho_{c}(H)$,~\cite{Eun:2011}  as   the magnetic field, electric field and the current are all in the same direction, and therefore normally there should not be  any oscillations of the two dimensional density of states; the picture presented there was  that the electron executes many cyclotron orbits in the plane and infrequently incoherently hops betweenn the planes. Second order  perturbation theory, assuming coherence, in terms of $\left(t_{\perp}/E_{F}\right)$ must  fail because of the degeneracy of the electronic states of the stacked copper-oxide planes, which leads to vanishing energy denominators. An exact analysis of the {\em density of states}  with a coherent energy dispersion in a simplified model was discussed by Stephen~\cite{Stephen:1992}  as  to how the two-dimensional quantum oscillations can be inherited in  $\rho_{c}(H)$, but unfortunately this model is totally inconsistent with the incoherent $c$-axis transport in underdoped cuprates.

In my opinion, the true explanation of these remarkable Kerr measurements and their implications remain unexplained.

This work was supported by  by NSF under Grant No. DMR-1004520.

\end{document}